# Voltage-Regulated Photoluminescence Modulation in a 0D-2D Mixed Dimensional Heterostructure

S. V. U. Vedhanth[1], Amit Bhunia[1], Mohit Kumar Singh[1], Yuvraj Chaudhry[1], Mohamed Henini[2] and Shouvik Datta[1,*]

[1]Department of Physics, Indian Institute of Science Education and Research, Pune 411008, Maharashtra, India

[2]School of Physics and Astronomy, University of Nottingham, Nottingham NG7 2RD, UK,

*Corresponding author: shouvik@iiserpune.ac.in.

**ABSTRACT**

Bias-dependent oscillations in excitonic photoluminescence are observed in a mixed-dimensional 0D–2D heterostructure. These oscillations arise from modulation by oscillatory DC photocurrent, which exhibits periodic negative differential resistance, indicating recurring charge accumulation within the heterostructure. The persistence of these oscillations across a macroscopic area of diameter ~200 $\mu$m suggests the presence of periodically correlated quantum phenomena over large length scales. Furthermore, bias-dependent oscillations in the photo-capacitance are observed, reflecting a periodic ordering and disordering of excitonic populations. Together, these observations point to a direct competition between coherent and incoherent electron tunnelling processes. The coupled oscillatory behaviour of photoluminescence, photocurrent, and photo-capacitance highlights new opportunities for exciton-based quantum optoelectronic devices.

**Key Words:** Resonant Tunnelling Diode, Quantum Dots, Exciton, Negative Differential Resistance, Photoluminescence

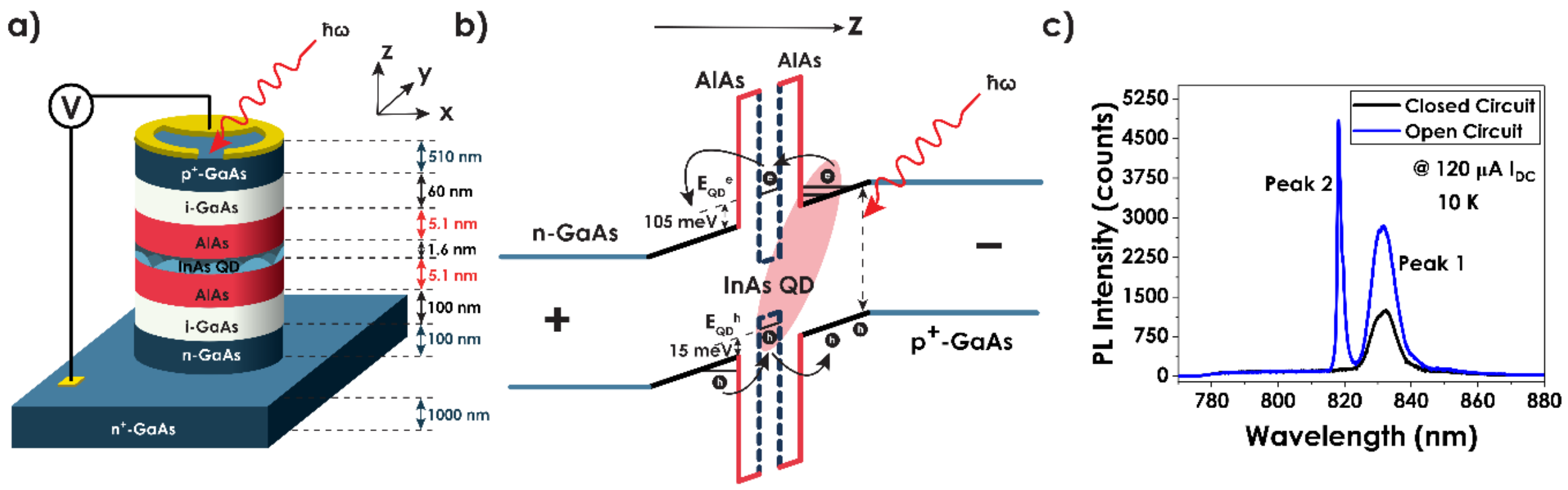

**Figure 1.** (a) A schematic representation of the p$^+$-GaAs/i-GaAs/AlAs/InAs QDs/AlAs/i-GaAs/n-GaAs based single crystal of a 0D-2D, mixed-dimensional heterostructure [17] and its (b) approximate energy band diagram along the z-direction under reverse bias V. Schematic figures are not proportional to its original size/thickness/width. This RTD has an array of InAs QDs sandwiched between two 5 nm thick AlAs potential barriers from both sides in the x-y plane of the sample. Here, the 0D refers to zero-dimensional InAs QD and 2D refers to the two-dimensional electron gas (2DEG) which is photogenerated and accumulated under applied reverse bias near the top p$^+$-GaAs side of the AlAs barrier. The shaded pair of electron and hole shows the 0D-2D spatially indirect exciton [17]. Average spatial extent of these InAs QDs before growing the capping layers were measured using Atomic Force Microscope (AFM). These are $\left(L^{Q}_{x,y}\right)$ ~11 nm and $\left(L^{Q}_{z}\right)$ ~1.6 nm. AC/DC impedance under photo excitations are measured across the top and bottom contacts of the mesa structure. (c) Photoluminescence spectra of the sample in closed-circuit and open-circuit configurations are measured at 10 K with excitation intensity corresponding to 120µA DC-photocurrent at 0V applied bias. The observed peak 1 is from the excitons in GaAs and the peak 2 is from the InAs QDs. Further details of sample growth, measurements and spectral analyses of PL can be found in the Supplementary Materials.

## 1. Introduction

III-V semiconductors are known for their wide applications in the field of optoelectronics devices. One such device is a Resonant Tunnelling Diode (RTD) [5-9]. Quantum tunnelling [1-4] in RTDs typically manifests as a peak current in its current-voltage characteristics, followed by a decline with increasing bias voltage – a hallmark of negative differential resistance (NDR). NDR based devices often finds its application in memory [10] and neuromorphic computing [11]. Oscillations of such NDR were investigated [12] in III-V heterostructures including lateral superlattices in the growth plane [13] with tunneling thickness ($L_{Tunnel}$) comparable with the thermal de Broglie wavelength ($\lambda_{DB}$) of electrons and having lateral periodicity lesser than the mean free path ($\lambda_{MFP}$) of electrons ($\gtrsim$ several hundreds of nm at 10 K). This can split the energy-momentum (E-k) bands into smaller Brillouin zones and confine electrons to produce such oscillatory NDR [14].

Beyond its well-known negative differential resistance (NDR) characteristics, the photoluminescence properties of such devices have received relatively little attention [15,16]. In this work, we investigate both the photoluminescence and photocurrent behaviour of one such RTD, incorporating self-assembled InAs quantum dots that introduce potential modulations in the x–y plane, as illustrated in figure 1. These 0D-2D heterostructures are known for producing oscillations of photo generated capacitance ($C^{Ph}$) and DC-photocurrent ($I^{Ph}_{DC}$) as a function of V [17-20]. Notably, these oscillations are totally absent in the dark. While previous studies, including ours, extensively utilize AC photoconductance ($G_{AC}$) and Capacitance-Voltage spectroscopy [17,21-23] under photoexcitation, detailed investigations of the PL of such quantum structures remained unexplored.

## 2. Experiment

The sample used for this study has an array of InAs Quantum dots sandwiched between two AlAs potential barriers [17]. This combined structure is embedded in the intrinsic region of the p-i-n diode made of GaAs. The schematic illustration of the sample and its band structure is shown in figure 1(b). Individual epilayers were grown by Molecular Beam Epitaxy (MBE) on a highly doped (100) $n^+$ GaAs substrate. The p-type and n-type layers were doped using Be and Si respectively. Initially, a 1.0 μm thick heavily n-doped (Si = $4 \times 10^{18}$ /cm$^3$) GaAs buffer layer was grown at a temperature of 600 °C at a growth rate of 1 $\mu$m/hour. A 100 nm thick n-doped (Si = $2 \times 10^{16}/cm^3$) GaAs layer, then a 100 nm thick undoped GaAs spacer layer, and then two intrinsic AlAs quantum barriers of thicknesses of 5.1 nm along with a 1.8 monolayer thick InAs QDs having areal density $\sim 1 \times 10^{11}/cm^2$ in between these AlAs barrier layers. From Atomic Force Microscope (AFM), average spatial (x-y) extent of these QDs is measured to be $11 \pm 2$ nm and the average z dimension are ~1.6 nm[1]. Then another 60 nm undoped GaAs layer was grown on top of AlAs. At last, a 0.51 μm thick heavily p-doped (Be = $2 \times 10^{18}/cm^3$) GaAs layer was deposited on top. The InAs QDs were grown at a temperature of 520 °C using the well-known Stranski–Krastanov growth mode. For the two-terminal based photo-current measurements described in this paper, a circular mesa of diameter 200 $\mu$m was fabricated. Ohmic contacts were prepared with Au/Ge/Ni alloying on the n+ GaAs substrate and with Au-Zn alloying for the top p+ GaAs layer of the heterostructure. Sharp PL peaks of InAs QDs under open circuit conditions [peak 2 in Fig. 1(c) and also in Supplementary Figures 2(a), 2(b) and 6(c)] indicate the monodisperse nature of these InAs QDs. However, this PL from InAS QDs vanishes as soon as the sample is connected to a closed-circuit, even without any applied bias. We think it is because of the fact that whenever there is a channel for the flow of charge carriers in QDs, the lighter electrons tunnel through the barriers more readily leading to vanishing of radiative recombination of carriers responsible for the PL emission from these QDs. Any further measurements presented below are carried out only in closed-circuit conditions.

Top contact metallization was ring-shaped with a diameter around 200 μm and ring thickness around 25 μm. This allowed normal incidence photoexcitation and subsequent collection of photoluminescence from the sample. These photoluminescence measurements were carried out by keeping the sample on a customized copper holder inside a closed-cycle cryostat CS-204S-DMX-20 procured from Advance Research Systems. The temperature of measurement is controlled with a Lakeshore (Model 340) temperature controller. The sample was illuminated from the top p-GaAs side and the photo-luminescence from the sample is also collected at 90° configuration. The photo-luminescence spectra are measured using Horiba iHR 320 Spectrometer with the grating of 600 groves/mm which gives spectral resolution of ~0.17 nm in case of our experimental setup. The signal is measured using Synapse EMCCD (Electron Multiplying Charge Coupled Device) with 1600 × 200 pixels each having dimension of 16 × 16 $\mu$m. Upon measurement of the spectra, it is corrected for the response function of the EMCCD and the grating. The DC Photo-current is monitored using Agilent's E4980A LCR Meter. A Melles-Griot He-Ne laser of 632.8 nm (maximum intensity of 35 mW) is used as photo excitation for PL measurements and a white LED light with CMOS camera is used for imaging the sample, with light spots nearly filling the interiors of that ring-shaped top electrical contact having diameter ~ 200 μm. To detect PL oscillations having sufficiently large signal-to-noise ratio, we use a photo excitation intensity ($I$) corresponding to 7$\mu$A of $|I_{DC}^{Ph}|$ at V=0 based on linear relationship measured between $|I_{DC}^{Ph}|$ & $I$ [17].

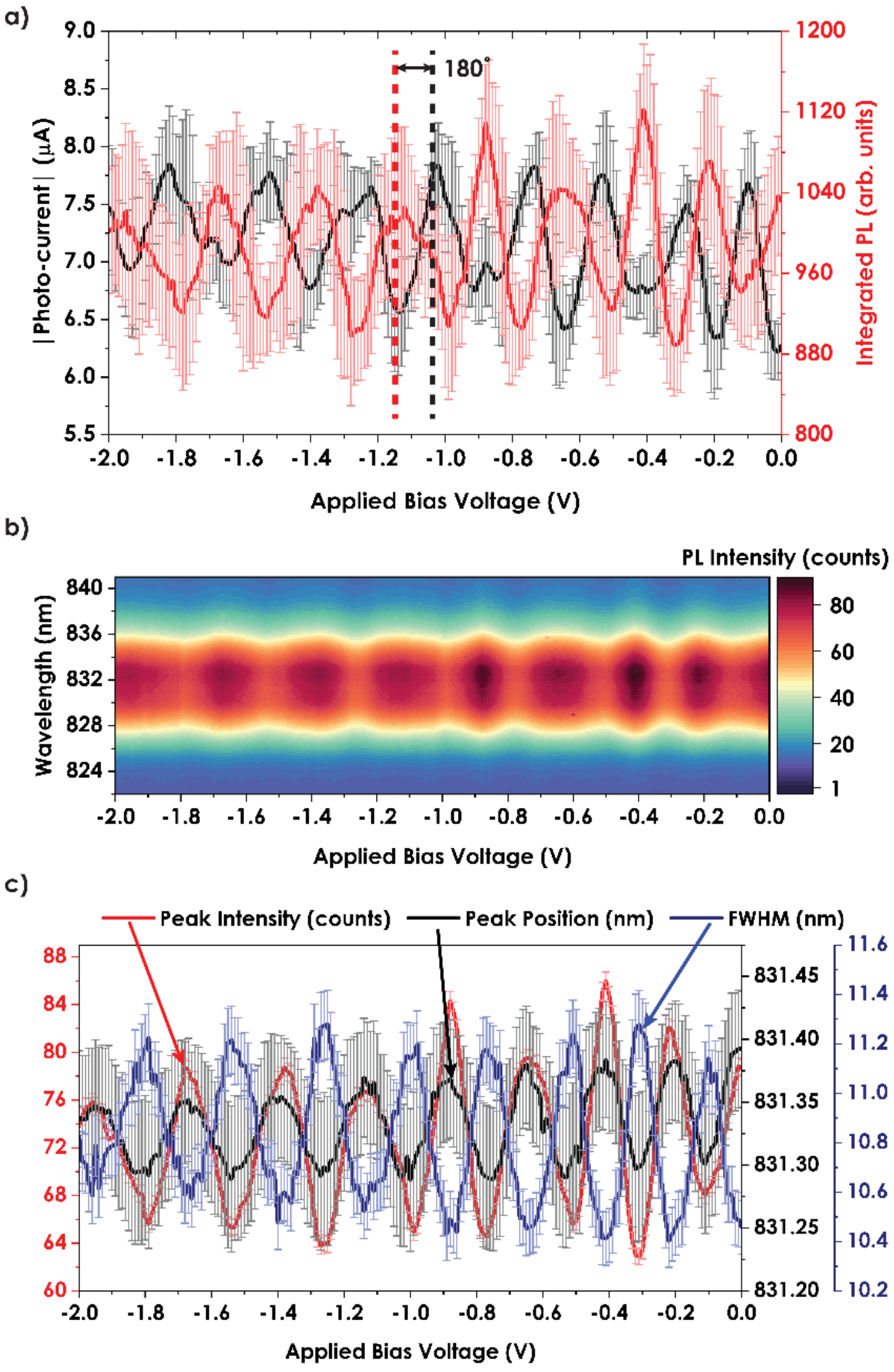

**Figure 2.** (a) Plot shows the observed oscillation of integrated PL intensity with bias and its 180° out-of-phase relationship with the absolute magnitude of $I_{DC}^{Ph}$ oscillation. More dc photocurrent through the 0D-2D heterostructure causes lesser number of available electron-hole pairs for PL from the GaAs 2DEG. Figure (b) Shows the contour plot of same oscillation of PL spectra with varying V. Interestingly, one can see that the width of the PL spectra clearly oscillates with V. Figure (c) shows the observed oscillations of Full-width-at-half-maxima (FWHM), peak position along with the oscillation of integrated peak intensity as a function of V. These are obtained from Gaussian Fitting of PL spectra. These details can be found in the Supplementary Materials.

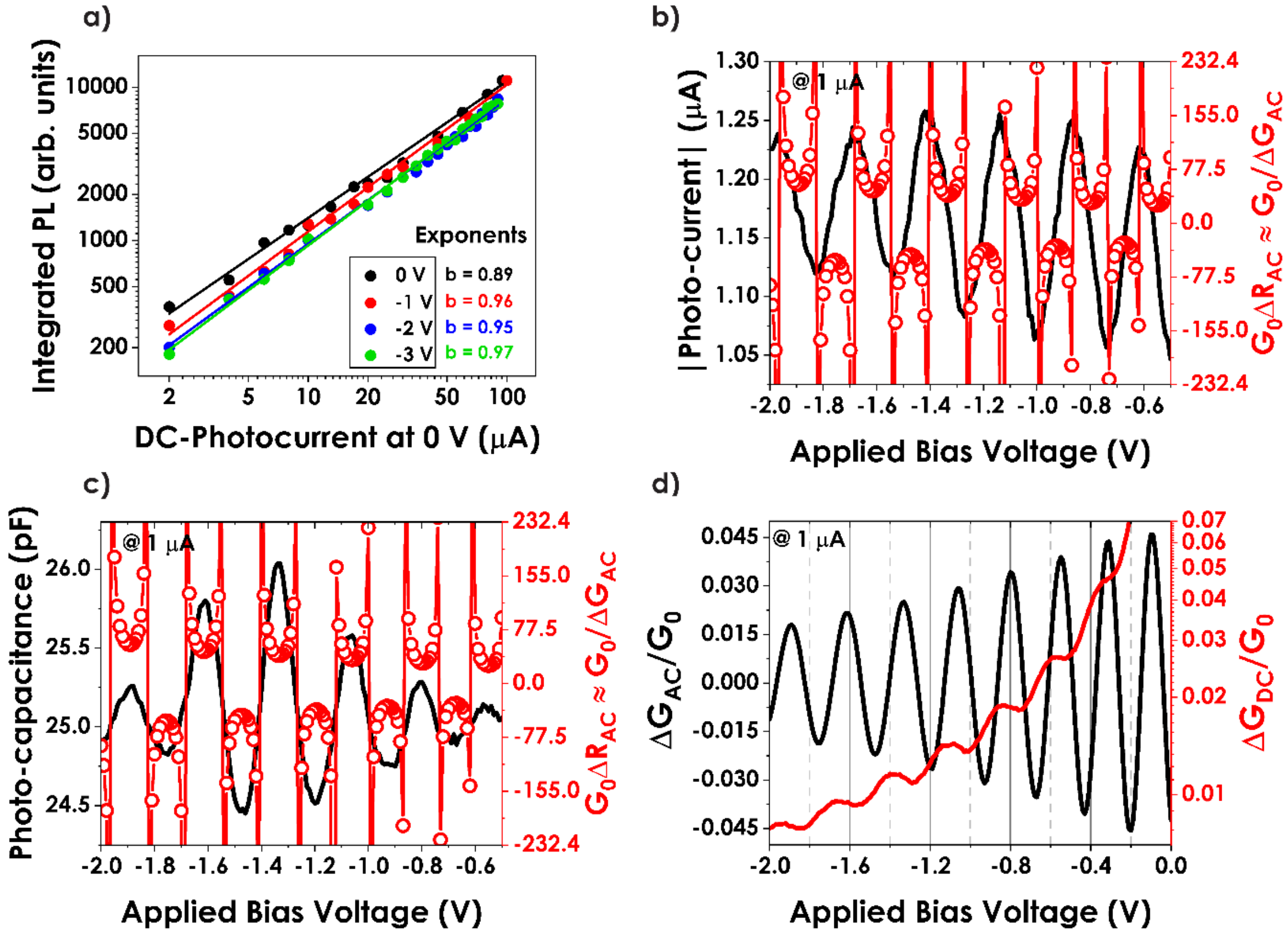

**Figure 3.** Plot (a) presents the power-law analyses of the integrated PL spectra under varying photo excitation intensity ($I$) by nearly two orders of magnitude. As mentioned in the text, we monitor $I$ by measuring $|I_{DC}^{Ph}|$ at V=0, such that $I$ is $\propto |I_{DC}^{Ph}|$. Observation of exponents close to unity across different bias voltages suggest that, despite the oscillations in PL with changing bias, the PL maintains its excitonic nature. (b) Shows $|I_{DC}^{Ph}|$ oscillation in comparison with $G_0\Delta R_{AC}$ as $\approx G_0/\Delta G_{AC}$ with $G_0 = \frac{2e^2}{h}$ and the oscillatory NDR with increasing V. Figure (c) shows the same $G_0\Delta R_{AC}$ along with $C^{Ph}$ oscillations for comparison. Figure (d) shows the $\Delta G_{AC/DC}$ as functions of reverse bias V. No error bars are added here for sake of clarity in comparisons. Photo excitation intensity ($I$) used in the last three plots corresponds to $|I_{DC}^{Ph}|$=1 μA at V=0.

### 3. Results

The figures 2(a) and 2(b), show the integrated intensity of the PL spectra (the peak 1 observed from GaAs in case of closed circuit as shown in figure 1(c)) oscillates and this oscillation is 180° out-of-phase with the $|I_{DC}^{Ph}|$ oscillation as a function of V. Both of these oscillations in PL and $|I_{DC}^{Ph}|$ are riding on some finite background value. Interestingly, the peak position and the FWHM of PL also oscillate with V [figures 2(b), 2(c)]. For each decrease in integrated PL intensity and accompanied increase in $|I_{DC}^{Ph}|$, the PL linewidth (FWHM) broaden and PL peak position shift toward higher energies. Integrated PL peak intensity [figure 3(a)] follow a power law with exponent one as a function of $I$ $(\propto |I_{DC}^{Ph}|)$ indicating their excitonic [25] origins. Whenever $I_{DC}^{Ph}$ decreases, the peak position of these excitonic PL shifts by ~0.3 meV to lower energies.

Although, presence of coherent resonant tunneling and macroscopically large, coherent state of photo generated and bias driven indirect excitons (IXs) at the 0D-2D heterojunction (figure 1) was inferred using $C^{Ph}$ in the past [17]. However, we did not observe any measurable optical interference pattern and large-scale spatial coherence of the PL measured using [26] unpolarized photo excitations at normal incidence. This is because $C^{Ph}$ only probes the collective electrical polarization of the dipoles of IXs which are optically dark [27] in these III-V materials. However, PL probes all excitons which radiatively recombine. PL spectra are observed even at V = 0, however, excitonic peaks were absent [17] from $C^{Ph}$ spectra at zero bias because there was no significant formation of any bias driven IXs.

Photo induced changes in measured conductance as $\Delta G_{AC/DC} = G_{AC/DC}^{ph} - G_{AC/DC}^{dark}$ are estimated and used in figures 3(b)-3(d), where the $G_{AC}^{dark} \sim 0$ and the $|G_{DC}^{dark}| << |G_{DC}^{ph}|$. Here the superscript 'dark' refers to no-illumination and 'ph' refers to illumination with 632.8 nm. Comparisons of raw data of $G_{AC/DC}^{ph}$ & $G_{AC/DC}^{dark}$ can be found at the end of Supplementary Materials. Figure 3(b) shows the differential AC resistance as $G_0\Delta R_{AC}$ $\left(\approx \frac{G_0}{\Delta G_{AC}}\right)$ vs V in units of $\frac{1}{G_0}$ where $G_0$ is the quanta of conductance ($G_0 = \frac{2e^2}{h}$) using 30 mV AC rms at 10 kHz under illumination. It reveals periodic appearance of NDR with increasing V. We witness a Peak-to-Valley-Current-Ratio $\left(PVCR = \frac{I_{peak}}{I_{valley}}\right)$ of NDR ~1.14 instead of the required PVCR~5 reported [13] earlier.

In figure 3(c), we see that NDR regions coincide with the minima of $C^{Ph}$ oscillations. The observed $\Delta G_{AC}$ was equivalent to $\frac{d(I_{DC}^{Ph})}{dV}$ [17]. So, whenever $I_{DC}^{Ph}$ reaches the maximum and begins to decrease, NDR starts to build up. Similarly, whenever the $I_{DC}^{Ph}$ reaches the minimum and begins to increase again, $\Delta R_{AC}$ changes sign and becomes positive. Figure 3(d) shows the comparison of $\Delta G_{AC}$ and $\Delta G_{DC}$ in units of $G_0$. As shown in figure 1, different overlapping population of excitons can contribute to PL, $|I_{DC}^{Ph}|$ and to $C^{Ph}$, which are not necessarily mutually exclusive. Dynamics between different populations of excitons in RTDs was explored earlier [28]. Usually, a single NDR region is associated with reduction of current and subsequent accumulation of electrons [29]. So, observation of oscillatory NDR regions in figures 3(b), 3(c) indicate periodic accumulation of electrons in the 2DEG with increasing V as will be discussed later in the context of figure 4.

**4. Discussion**

Moreover, the observed *'fractional'* nature of photogenerated variations of $\Delta G_{AC}$ and $\Delta G_{DC}$ in the units of $G_0$ in figures 3(c), 3(d) indicate the existence of few ingredients – (i) presence of interacting, correlated [30] electrons inside the 2DEG and (ii) inhomogeneities [31] in these nanoscale conducting channels. These issues are usually discussed in the context of fractional quantum Hall effects and quantum point contact (QPC). However, we have not applied any magnetic field here. The observed features in $\Delta G_{AC/DC}$ are oscillatory rather than peak like [figure 3(d)]. However, periodic changes in the FWHM of PL with V [figure 2(b), 2(c)] can be a direct indicator of presence of periodic changes in order or disorder of the 2DEG electrons related to excitonic populations which recombine radiatively. We had reported [17] the existence of such interacting, Coulomb correlated electrons and associated presence of excitonic order as Bose-Einstein Condensate (BEC) coinciding with maxima of $C^{Ph}$ in this particular 0D-2D heterostructure with exciton densities reaching ~$10^{11}/cm^2$. Secondly, physical size distribution and placement of MBE grown QDs in the x-y plane won't change significantly with increasing bias along z. So, we ignore inhomogeneous broadening from size distributions of the InAs QDs and also from coupling to electrical contacts in the following analyses for simplification only. Interestingly, the presence of strongly confined InAs QDs (assuming $\lambda_{DB}$ of InAs as ~40 nm >> $L^{Q}_{x,y}$) can provide *'multiple'*, parallel, coherent, resonant tunneling channels for electron waves as a 2D array of QPCs. Here $L_{Tunnel}$ ~ << $\lambda_{MFP}$ and estimated number of occupied quantum modes of electron are $\left(N \approx 2L^{Q}_{x,y}/\lambda_F\right) \lesssim 1$, assuming the Fermi Wavelength ($\lambda_F$) ~35 nm for InAs. Therefore, large quantum uncertainties of electron momentum within these strongly confined InAs QDs can lead to robust 'pinched-off regions' [32] for ballistic conduction of these 2DEG electrons and thereby produce the observed *'fractional'* quantum conductance oscillations in both ac and dc photoconductance in figure 3(d). However, such collective nature of ballistic conductions through ~millions of InAs QDs acting as QPCs must be 'synchronized' within a macroscopically large region in the x-y plane, so that phase-coherent interference of electron waves can produce the oscillatory quantum conductance. Otherwise, electron waves transmitting through all these QDs can add up incoherently and observed oscillations could easily get wiped out. This is all the more important when we are not using any 'single point-contact' based measurements. We had already discussed [17] why such quantum coherence survives even in presence of tunneling induced decoherence in this 0D-2D heterostructure. Specifically, the following observations [17] had demonstrated robust presence of long-range coherence - (a) larger amplitude oscillations under forward bias with smaller periods, whenever more massive holes are taking part in the tunneling process as compared to electrons tunneling under reverse bias, (b) the critical role of in-plane coulomb correlation in 2DEG as evidenced from negative quantum capacitance (NQC), (c) generation of interference of oscillations of $C^{Ph}$ when excited with two overlapping light spots etc. So, this requirement of phase-coherent, synchronized tunneling of electron waves definitely need long-range spatial coherence spread over many QDs [33]. This was linked [17] with the narrowing of the distribution in momentum space of all these 2DEG electrons in the x-y plane, so that these 2DEG electrons can periodically satisfy the resonant tunneling through QDs as given below,

$$E^{e}_{QD}(V, I) = E^{e}_{2DEG}(V, I) + \left[\frac{\hbar^2}{2m^*}\left(\vec{k}_x^2 + \vec{k}_y^2\right) + \Phi^{EX}_{xy}(V, I)\right] + \mathcal{O}(V) \quad (1)$$

where $I$ is the light intensity and V is applied bias, $E^{e}_{QD}(V, I)$ and $E^{e}_{2DEG}(V, I)$ are the ground state energy levels of electrons in QD and 2DEG respectively, $\vec{k}_x$ and $\vec{k}_y$ are in-plane momentum of 2DEG electrons, $\Phi^{EX}_{xy}(V, I)$ is the in-plane coulomb interaction (correlation) energy of these 0D-2D IXs, which was associated with the observed NQC and BEC of excitonic dipoles [17], $\mathcal{O}(V)$ can be any other voltage-dependent changes in energies which are negligible. Notably, the terms within the square bracket in equation (1) have to be same for all electrons taking part in resonant tunneling $\left[E^{e}_{QD} - E^{e}_{2DEG} = 0\right]$. Such resonant tunneling of electrons between 2DEG and QDs and the associated excitonic BEC occur at the maxima of $C^{Ph}$, where the collective average of electrical polarization of quantum clones of excitonic dipoles over a macroscopically large area maximize in the ground state of a BEC and orient ~identically along z [17]. Consequently, it is expected from figure 3(c) that the NDR oscillations are intertwined with $C^{Ph}$ and connected with spatially correlated, interacting 2DEG electrons and associated itinerant presence and absence of BEC of excitonic dipoles as a function of applied bias over a macroscopically large area.

Next in figure 4(a), we describe this proposed mechanism where either phase-coherent tunneling of 2DEG electrons through InAs QDs dominates or incoherent tunneling with phase randomization governs the physics. Whenever $E^{e}_{2DEG}(V, I)$ moves out of resonance with $E^{e}_{QD}(V, I)$, it leads to increase in the phase randomization of the 2DEG electrons as the coherent resonant tunneling condition and k-space narrowing dictated by equation (1) is not met anymore. So, this also increases the incoherent tunneling probability as compared coherent tunneling [33] at resonance. It was known [33] that while transitioning from coherent to incoherent tunneling regime, it increases off-resonant tunneling in the crossover region and this would also be accompanied by increase in accumulation of carriers. As mentioned above [17], the maxima of $C^{Ph}$ (and $\Delta G_{AC}$) lies in the coherent regime of tunneling and the corresponding minima lies in the incoherent regime. So, the observation [figure 3(b)] of $\left|I_{DC}^{Ph}\right|$ maxima/minima at in-between those biases indicate the delayed changes in $E_F$ with respect to $E_{2DEG}$ as a result of increase in carriers with increase/decrease in the off-resonance tunneling probability once the system is being driven away-from/toward resonance. Increased coulomb screening effects at maximum $\left|I_{DC}^{Ph}\right|$ lowers the binding energy of 2DEG excitons and coincides with a measurable increase in PL peak energy here as shown in figure 2(c). Upon

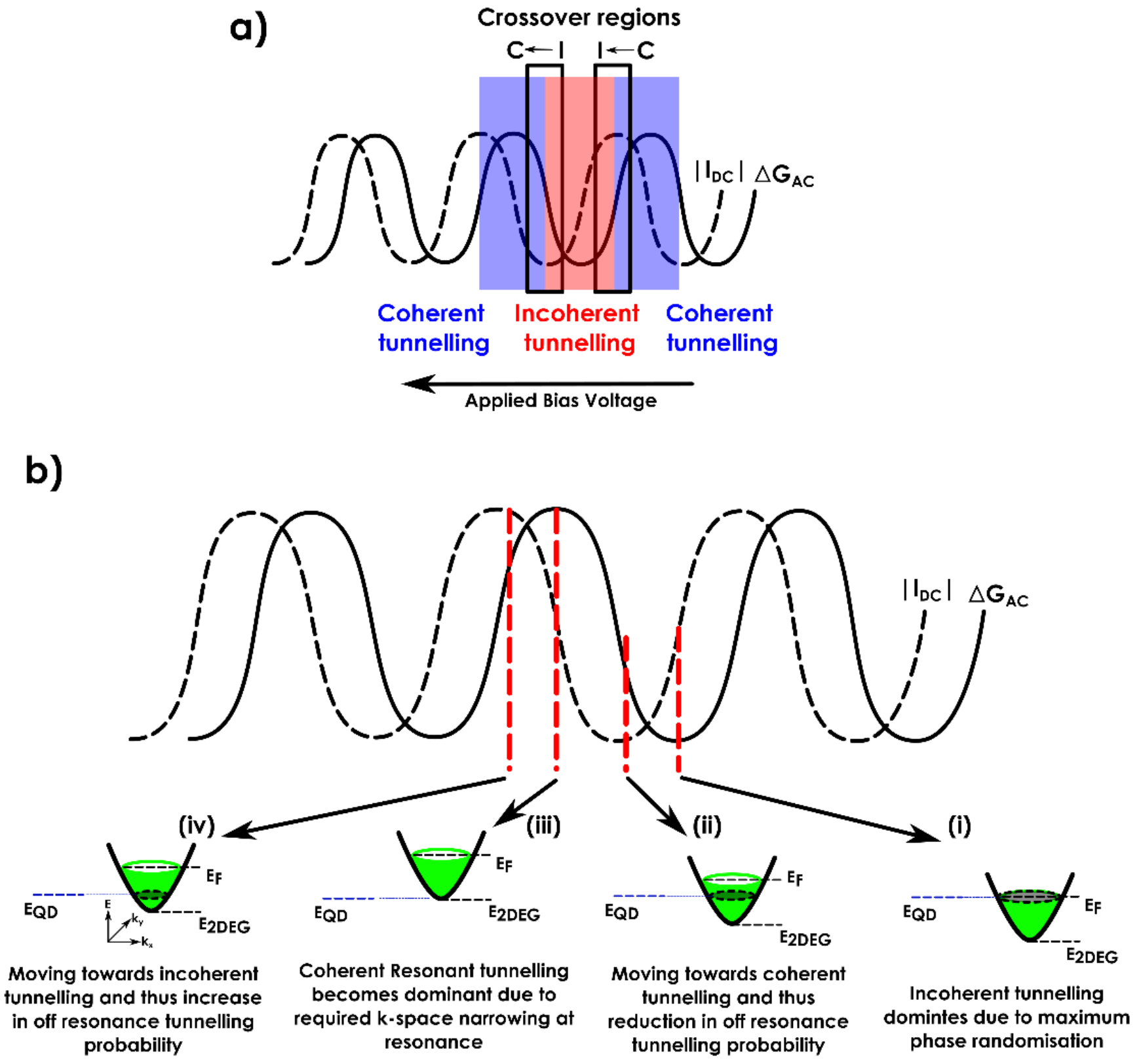

**Figure 4.** The schematic (a) presents the rationale behind the observed oscillations of PL, $\left|I_{DC}^{Ph}\right|$ and $G_{AC}$ at different bias regimes where the tunnelling is either dominated by its coherent (C) or incoherent (I) parts and their dynamic competitions to explain results in figures 2, 3 as discussed in the text. The boxed areas represent the crossover regions from coherent to incoherent tunnelling regions and vice versa. Following the equation (1), the schematic (b) shows the periodic changes in E-k distribution of 2DEG electrons which govern the coherent/incoherent tunnelling fractions as we go from (i) to (iv) with increasing bias. $E_F$ is the Fermi Energy of 2DEG. Red dashed lines correspond to related coherent/incoherent portions of the oscillatory $\Delta G_{AC}$ and $\left|I_{DC}^{Ph}\right|$.

crossing into incoherent regime, the $\left|I_{DC}^{Ph}\right|$ decreases as shown in (i) of figure 4(b). This generates more accumulation of 2DEG electrons leading to corresponding increase in PL [figure 2(a)] and the onset of NDR is also observed [figure 3(b)]. Then $\left|I_{DC}^{Ph}\right|$ reaches a minimum as in (ii) of figure 4(b) and PL maximize [figure 2(a)]. Thereafter, the off-resonance tunneling probability starts to decrease with increasing V and $E_{2DEG}$ finally moves towards the resonant condition $\left[E_{QD}^{e} - E_{2DEG}^{e} = 0\right]$ within the coherent regime as depicted in (iii) of figure 4(b).

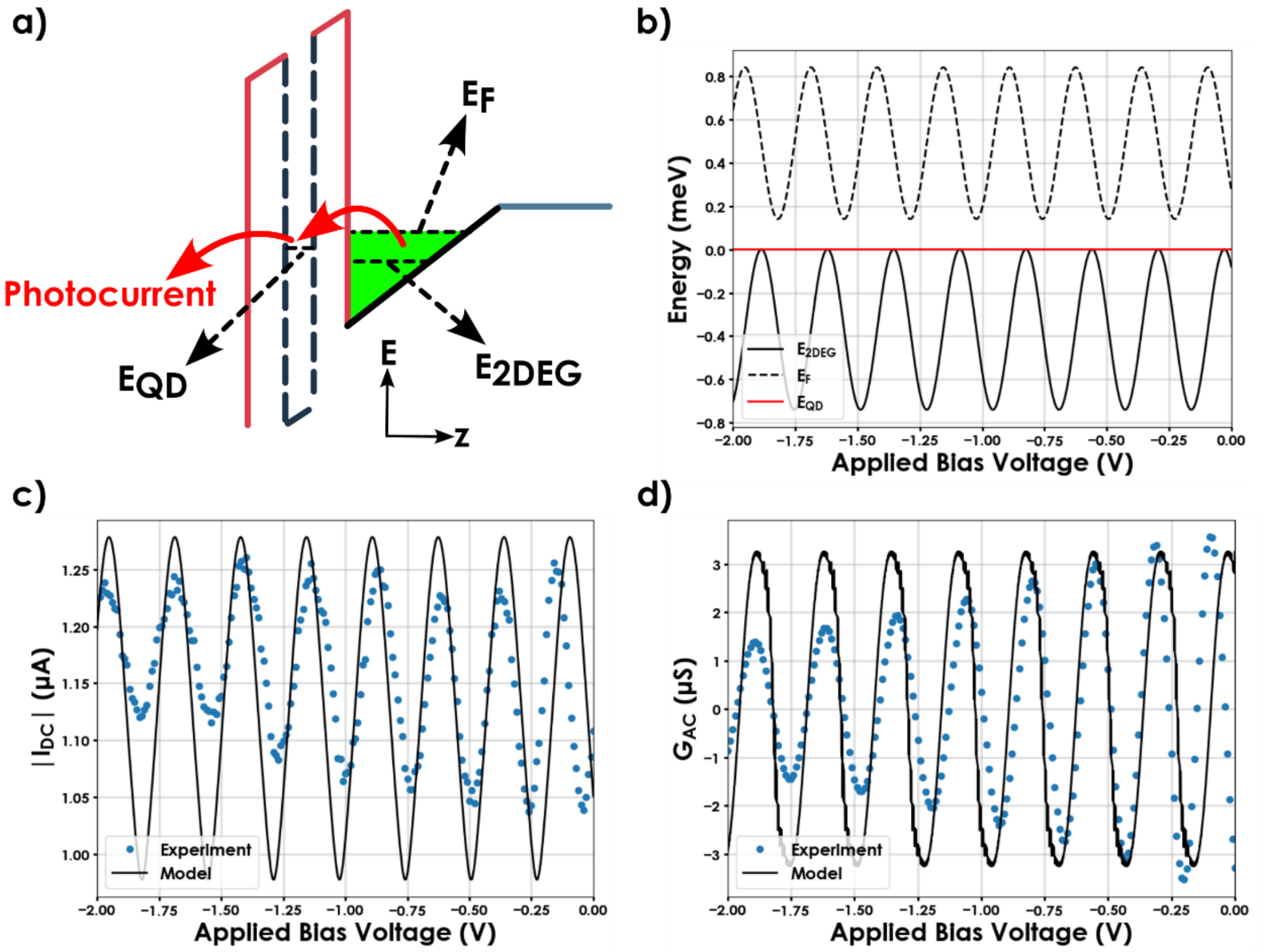

**Figure 5.** The figure (a) shows the illustration of the directionality of the photocurrent calculated in the simulation. The figure (b) shows the oscillating energy levels $E_{2DEG}$ ($E_{2DEG} = \left(A_{2DEG} \times sin\left(\frac{2\pi V}{period} + \varphi\right)\right) - A_{2DEG}$ ) and $E_F$ ($E_F = (1.4 \cdot A_F) + \left(A_F \times cos\left(\frac{2\pi V}{period} + \varphi\right)\right)$ ) with respect to $E_{QD}$ which is taken as 0. The figure (c) and (d) shows the calculated photocurrent and AC-conductance. The photocurrent is given by equation $(A \cdot I_{DC}) + C$ where $I_{DC}$ is the calculated DC-photocurrent from equation and is scaled by the parameters A and C. The $I_{DC}$ is calculated as shown in equation (2) and $G_{AC} = \frac{d}{dV} I_{DC}$. The parameter values chosen for the simulation plot are $k_B T = 0.86\, meV\ (at\ 10K)$, $\Gamma = 0.1\, meV$, $A_{2DEG} = 0.37\, meV$, $A_F = 0.35\, meV$, $period = 0.3\, V$, $T_{ref} = 1.7\, \mu A$, $\varphi = 2.24\, rad$, $A = 4.5$ and $C = -0.68\, \mu A$.

Upon further increase in bias, $|I_{DC}^{Ph}|$ moves again towards a maximum and lesser electron accumulation in 2DEG lowers $E_{2DEG}$. Consequently, the resonant condition is lifted as shown in (iv) of figure 4(b). These cycles between coherent and incoherent tunneling processes then repeat over and over with increasing V. Using figure 4(b) and equation (1), we perceive that oscillation periods of both $|I_{DC}^{Ph}|$ and PL reflect the difference of [$E_F$ - $E_{2DEG}^{e}$], which also strongly depend on $\Phi_{xy}^{EX}(V, I)$ as shown in the past [17]. Accumulation of excess electrons at the 0D-2D junction increases under increasing V. So, the incoherent part of the background leakage current, over and above which the coherent $I_{DC}^{Ph}$ oscillations are riding [figures 2(a), 3(b)], actually increases with increasing reverse bias and gradually suppress the phase-coherent oscillation amplitudes of $\Delta G_{AC}/G_0$ and also the step sizes of $\Delta G_{DC}/G_0$ [figure 3(d)].

To simulate the oscillations in $\left|I_{DC}^{Ph}\right|$ and $\Delta G_{AC}$ measurements, we calculate at the resonant tunnelling current in case of 2DEG across a barrier. The resonant tunnelling current in case of a 2DEG, where potential vary only across one axis, across a barrier is given as

$$I_{res} \propto J_{res} = \frac{-e}{h} \int_{\max(E_{2DEG}, E_{QD})}^{E_F} n_{2D}(E_F - E) \cdot T(E) dE \quad (2)$$

and the sign indicates directionality shown in the figure 5. The current from other direction is not considered as the device is reverse biased. Here $n_{2D}$ is the density of states in a sub-band given as $n_{2D}(E_F) = \frac{mk_BT}{\pi\hbar^2}\ln(1 + e^{E_F/k_BT})$ and $T(E)$ is the transmission coefficient for a resonant phenomenon given as $T(E) \propto T_{res}\left[1 + \left(\frac{E-E_{res}}{\frac{1}{2}\Gamma}\right)^2\right]^{-1}$ [34]. To simulate the observed oscillations both E2DEG and EF are made to oscillate to simulate the changing well depth due to accumulation and discharge of carriers with applied bias. EF is made to oscillate with a phase shift of 90° to account for the delay in its oscillation caused by combined effect of increase in accumulation(discharge) of carriers and increase(decrease) in the off-resonance tunnelling probability upon crossing from coherent tunnelling regime to incoherent tunnelling regime (vice versa). The energy levels considered here are shown in figure 5(b) and the calculated photo-current and AC-photoconductance, which is derivative of photo-current in our experiment, are shown in figure 5(c) and 5(d) respectively. The values of choice for the parameters are not from the fit but rather handpicked to match the experimental results. This employed model is just a simple model explaining that the argument of oscillation of energy levels is required to explain the observed oscillation in DC-photocurrent. This model also doesn't include the increase in decoherence effects that happens with increase in bias voltage which would lead to damping of observed oscillation. Nevertheless, this model shows that the oscillation is a result of energy levels changing periodically due to changes in accumulated charge density across the barriers and not from a sequential tunnelling from the states of the 2DEG.

The observed $\left|I_{DC}^{Ph}\right|$ oscillation, as a function of V, is different from the usual plateaus in the standard current-voltage (I-V) characteristics of the Coulomb Blockade [32, 35-37]. We observe these oscillations only when the sample is illuminated and these also doesn't show any oscillatory features in dark conditions as shown in the figure 7 of the Supplementary Materials. On top of that, these oscillations are observed for a collection of many quantum dots (~$10^{11}$/cm$^2$) rather than a single quantum dot used in standard QPC [32,38] based experiments.

Although each QD can act as a nanoscale capacitor, the effective large area capacitance (C) of our sample containing nearly ~$10^7$ InAs QDs at 10 K is ~25 pF upon no applied bias and without any photoexcitation. So, the charging energy of this large area sample under measurement is $e^2/2C \approx 3\times10^{-9}$ eV, which is lower than the thermal energy at 10 K (~ $86\times10^{-5}$ eV). Most importantly, the photo generated AC conductance changes sign across the dark value [figures. 3(c), 3(d)] which clearly differentiate it from Coulomb blockade [32, 35-38] where the conductance need not oscillate between positive and negative values. Moreover, the figure 1(c) clearly shows that the PL from InAs QDs vanish in the closed-circuit condition under which most of above measurements were carried out. Although, the observed $C^{Ph}$, $\Delta G_{AC}$ & $I_{DC}^{Ph}$, PL oscillations as a function of applied voltage bias are certainly influenced by a collective phenomenon over a macroscopically large area in the form of periodic occurrence of coherent resonant tunnelling at the 0D-2D heterojunction, as explained in figures 4, 5. However, excitons or electron-hole pairs photogenerated anywhere within the sample structure, especially in the GaAs 2DEG as well as in the top p-type GaAs layer can actually contribute to both $I_{DC}^{Ph}$ & PL. Because of these above reasons, we concluded that the physical origin of these observed oscillations cannot be described by any simple-minded Coulomb blockade model for one single QD measured using a point-contact experiment.

## 5. Conclusion

To summarize, we observed that the integrated intensity of PL oscillates with varying voltage bias, which is modulated by 180° out-of-phase with the measured $|I_{DC}^{Ph}|$ oscillations. This occurs because the coherent, resonant tunneling condition between the quantized levels of the 2DEG and QDs is periodically satisfied only at specific bias voltages as shown also by the simulation. Exciton assisted sequential resonant tunneling phenomena was studied [39]. However, excitonic presence in oscillatory PL, simultaneous occurrence of oscillatory both NDR and fractional quantum conductance [30] measured as a function of bias over such a macroscopically large area strongly indicate – (i) a dynamical competition between coherent and incoherent tunneling and (ii) spatially large quantum correlation. These observations validate our earlier study [17] on having a bias dependent, 'itinerant' BEC of dipolar excitons probed by photo-capacitance using this 0D-2D heterostructure.

QD based similar 0D-2D, double-barrier, resonant tunnelling diodes made with Transition Metal D-Chalcogenides (TMDC), Perovskites, Nitrides, Oxides etc. materials having higher excitonic binding energies as compared to III-Vs materials and with more ordered, periodic arrays of QDs in the growth plane can lead to observation of optically-induced quantum oscillations of NDR even at room temperatures and above, possibly with even better PVCR. This highlights the potential of these heterostructures as efficient optoelectronic modulators, switches, memory, oscillator circuits [40-42] to leverage the inherent quantum origins of these oscillations. Such voltage-controlled oscillations of PL between highest and lowest intensity, can be modeled as optical bits of 1 and 0 respectively. Hence, these experimental observations are important not only for understanding the collective, many-body condensed matter physics of excitons, but also for exciton-based quantum computations [17] and quantum information processing.

## Acknowledgement

SD acknowledges the Science and Engineering Board (SERB) of Department of Science and Technology (DST), India (Grants # DIA/2018/000029, CRG/2019/000412) for supports.

## Data availability statement

Currently the data is not publicly available online. However, all data used in this paper will be made available on request as .CSV files. Authors agree to make any data required to support or replicate claims made in an article available privately to the journal's editors, reviewers and readers without undue restriction or delay if requested.

## References

[1] Tsu R and Esaki L 1973 Tunnelling in a finite superlattice *Appl. Phys. Lett.* **22** 562

[2] Esaki L 1958 New Phenomenon in Narrow Germanium $p-n$ Junctions *Phys. Rev.* **109** 603

[3] Miyamoto K and Yamamoto H 1998 Resonant tunnelling in asymmetrical double-barrier structures under an applied electric field *J. Appl. Phys.* **84** 311

[4] Tao B, Wan C, Tang P, Feng J, Wei H, Wang X, Andrieu S, Yang H, Chshiev M, Devaux X, Hauet T, Montaigne F, Mangin S, Hehn M, Lacour D, Han X, and Lu Y 2019 Coherent Resonant Tunnelling through Double Metallic Quantum Well States *Nano Lett.* **19** 3019

[5] Kinoshita K, Moriya R, Okazaki S, Zhang Y, Masubuchi S, Watanabe K, Taniguchi T, Sasagawa T and Machida T 2022 Resonant Tunnelling between Quantized Subbands in van der Waals Double Quantum Well Structure Based on Few-Layer $WSe_2$ *Nano Lett.* **22** 4640

[6] Chang L L, Esaki L, Tsu R 1974 Resonant tunneling in semiconductor double barriers *Appl. Phys. Lett.* **24** 593

[7] Park K W, Kang S J, Ravindran S, Min J W, Lee S K, Park M S and Lee Y T 2015 Resonant tunneling in semiconductor double barriers *Appl. Phys. Express* **8** 062302

[8] Roy T, Tosun M, Cao X, Fang H, Lien D H, Zhao P, Chen Y Z, Chueh Y L, Guo J and Javey A 2015 Dual-Gated $MoS_2$/$WSe_2$ van der Waals Tunnel Diodes and Transistors *ACS Nano* **9** 2071

[9] Oehme M, Sarlija M, Hahnel D, Kaschel M, Werner J, Kasper E and Schulze J 2010 Very High Room-Temperature Peak-to-Valley Current Ratio in Si Esaki Tunneling Diodes *IEEE Transactions on Electron Devices* **57** 2857

[10] Chen Y T, Santiago S R M S, Sharma S, Wu C B, Chou C L, Chang S H, Chiu K C and Shen J L 2022 Resistive Switching Accompanied by Negative Differential Resistance in Cysteine-Functionalized $WS_2$ Quantum Dots toward Nonvolatile Memory Devices *ACS Appl. Nano Mater.* **5** 2250

[11] Pei Y, Yang B, Zhang X, He H, Sun Y, Zhao J, Chen P, Wang Z, Sun N, Liang S, Gu G, Liu Q, Li S and Yan X 2025 Ultra robust negative differential resistance memristor for hardware neuron circuit implementation *Nat. Commun.* **16** 48

[12] Esaki L and Tsu R 1970 Superlattice and Negative Differential Conductivity in Semiconductors *IBM J. Res. Dev.* **14** 61

[13] Sakaki H, Wagatsllma K, Hamasaki J and Saito S 1976 Possible applications of surface-corrugated quantum thin films to negative-resistance devices *Thin Solid Films* **36** 497

[14] Ismail K, Chu W, Antoniadis D A and Smith H I 1988 Surface-superlattice effects in a grating-gate GaAs/GaAlAs modulation doped field-effect transistor *Appl. Phys. Lett.* **52** 1071

[15] Roy S S, Aktar S, Tamang A, Biswas K and Chattopadhyay B 2026 0D/2D Nanomaterials Heterostructures for High-Performance Photodetectors: Combining Quantum Dots With 2D Materials *Small22* **14** e09786

[16] Boulesbaa A, Wang K, Samani M M, Tian M, Puretzky A A, Ivanov I, Rouleau C M, Xiao K, Sumpter B G and Geohegan D B 2016 Ultrafast Charge Transfer and Hybrid Exciton Formation in 2D/0D Heterostructures *J. Am. Chem. Soc.* **138** 14713

[17] Bhunia A, Singh M K, Huwayz M A, Henini M and Datta S 2023 0D-2D heterostructure for making very large quantum registers using 'itinerant' Bose-Einstein condensate of excitons *Materials Today Electronics* **4** 100039

[18] Vdovin E E, Ashdown M, Patane A, Eaves L, Campion R P, Khanin Y N, Henini M and Makarovsky O 2014 Quantum oscillations in the photocurrent of GaAs/AlAs p-i-n diodes *Phys. Rev. B* **89** 205305

[19] Belyaev A E, Eaves L, Main P C, Polimeni A, Stoddart S T and Henini M 1998 Capacitance Spectroscopy of single-barrier GaAs/AlAs/GaAs structures containing InAs quantum dots *Acta. Phys. Polonica A* **94** 245

[20] Vdovin E E and Khanin Y N 2021 Effect of the Radiation Power on the Modification of Oscillations of the Photocurrent in Single-Barrier p–i–n GaAs/AlAs/GaAs Heterostructures with InAs Quantum Dots *JETP Lett.* **113** 586

[21] Capasso F, Mohammed K and Cho A 1986 Resonant tunnelling through double barriers, perpendicular quantum transport phenomena in superlattices, and their device applications *IEEE Journal of Quantum Electronics* **22** 1853

[22] Bhunia A, Singh M K, Gobato Y G, Henini M and Datta S 2018 Experimental evidences of quantum confined 2D indirect excitons in single barrier GaAs/AlAs/GaAs heterostructure using photocapacitance at room temperature *J. Appl. Phys.* **123** 044305

[23] Pal S, Junggebauer C, Valentin S R, Eickelmann P, Scholz S, Ludwig A and Wieck A D 2016 Probing indirect exciton complexes in a quantum dot molecule via capacitance-voltage spectroscopy *Phys. Rev. B* **94** 245311

[24] Labud P A, Ludwig A, Wieck A D, Bester G and Reuter D 2014 Direct Quantitative Electrical Measurement of Many-Body Interactions in Exciton Complexes in InAs Quantum Dots *Phys. Rev. Lett.* **112** 046803

[25] Datta S and Marie X 2024 Excitons and excitonic materials *MRS Bulletin* **49** 1

[26] Vedhanth S V U and Datta S 2023 Direct determination of 2D momentum space from 2D spatial coherence of light using a modified Michelson interferometer *Rev. Sci. Instrum.* **94** 095110

[27] Combescot M, Matibet O B and Combescot R 2007 Bose-Einstein Condensation in Semiconductors: The Key Role of Dark Excitons *Phys. Rev. Lett.* **99** 176403

[28] Camps I, Makler S S, Vercik A, Gobato Y G, Marques G E and Brasil M J S P 2005 The dynamics of excitons and trions in resonant tunneling diodes *Solid State Communications* **135** 241

[29] Lin J and Ma D 2008 Origin of negative differential resistance and memory characteristics in organic devices based on tris(8-hydroxyquinoline) aluminum *J. Appl. Phys.* **103** 124505

[30] Costa M R D, Shelykh I A and Bagraev N T 2007 Fractional quantization of ballistic conductance in one-dimensional hole systems *Phys. Rev. B* **76** 201302

[31] Sanvito S, Kwon Y K, Tománek D and Lambert C J 2000 Fractional Quantum Conductance in Carbon Nanotubes *Phys. Rev. Lett.* **84** 1974

[32] Houten H and Beenakker C 1996 Quantum Point Contacts *Phys. Today.* **49** 22

[33] Buttiker M 1988 Coherent and sequential tunneling in series barriers *IBM J. Res. Dev.* **32** 63

[34] Davis J H 1998 *The Physics of Low Dimensional Semiconductors* (Cambridge University Press)

[35] Fulton T A and Dolan G J 1987 Observation of Single-Electron Charging Effects in Small Tunnel Junctions *Phys. Rev. Lett.* **59** 109

[36] Ding Z, Quinn B M, Haram S K, Pell L E, Korgel B A and Bard A J 2002 Electrochemistry and Electrogenerated Chemiluminescence from Silicon Nanocrystal Quantum Dots *Science* **296** 1293

[37] Mouafo L D N, Godel F, Simon L, Dappe Y J, Baaziz W, Noumbé U N, Lorchat E, Martin M B, Berciaud S, Doudin B, Ersen O, Dlubak B, Seneor P and Dayen J F 2021 0D/2D Heterostructures Vertical Single Electron Transistor *Adv. Funct. Mater.* **31** 2008255

[38] He J, Yue X and Guo H 2020 Mesoscopic capacitance oscillations due to quantum dynamic coherence in an interacting quantum capacitor *Appl. Phys. Lett.* **117** 113103

[39] Cao S M and Willander M 1997 Exciton-induced tunnelling effect on the current-voltage characteristics of resonant tunnelling diodes *J. Appl. Phys.* **81** 6221

[40] Boylestad R L and Nashelsky L 2011 *Electronic Devices and Circuit Theory 11th ed.* (Pearson)

[41] Horowitz P and Hill W 1959 *The Art of Electronics 2nd ed.* (Cambridge University Press)

[42] Sterzer F 1967 Tunnel Diode Devices *Advances in Microwaves* **2** 1

# ***Supplementary Materials***

# Voltage-Regulated Photoluminescence Modulation in a 0D-2D Mixed Dimensional Heterostructure

S. V. U. Vedhanth[1], Amit Bhunia[1], Mohit Kumar Singh[1], Yuvraj Chaudhury [1], Mohamed Henini[2] and Shouvik Datta[1,*]

[1]Department of Physics, Indian Institute of Science Education and Research, Pune 411008, Maharashtra, India

[2]School of Physics and Astronomy, University of Nottingham, Nottingham NG7 2RD, UK,

*Corresponding author: shouvik@iiserpune.ac.in.

**1. Analyses of Photoluminescence (PL) Data**

The excitons [1] or electron-hole pairs which contribute to $I_{DC}^{Ph}$ and PL measurements, are qualitatively different from those of the 0D-2D spatially IXs with large dipole moments which produced $C^{Ph}$ oscillations in the past [2]. Different populations of excitons are generated in this RTD heterostructure and all of these contribute differently to different measurements, leading to diverse manifestations of interconnected physical phenomena.

$$n_{total}^{Excitons} \longrightarrow n_{photocapacitance},\ n_{dc-photocurrent},\ n_{PL} \tag{1}$$

Here '$n$' represents the overlapping population of excitons, which are not necessarily mutually exclusive. Here the subscripts refer to the process where it contributes to either $C^{Ph}$ (it probes indirect excitons (IXs) formed across the AlAs barrier, which contribute to oscillations of collective electrical polarizations of excitonic dipoles) or $I_{DC}^{Ph}$ (it probes excitons that are generated anywhere in the sample which subsequently dissociate into free electrons and holes) or PL (it probes excitons that are generated anywhere in the sample, but subsequently recombine and emit light). Here we disregard the photo excited electrons, holes which are captured by electronic defects and not contributing to measured photocapacitance, dc-photocurrent and PL based on measurement frequency and temperature.

## 1.1 PL in case of Closed-Circuit Measurements:

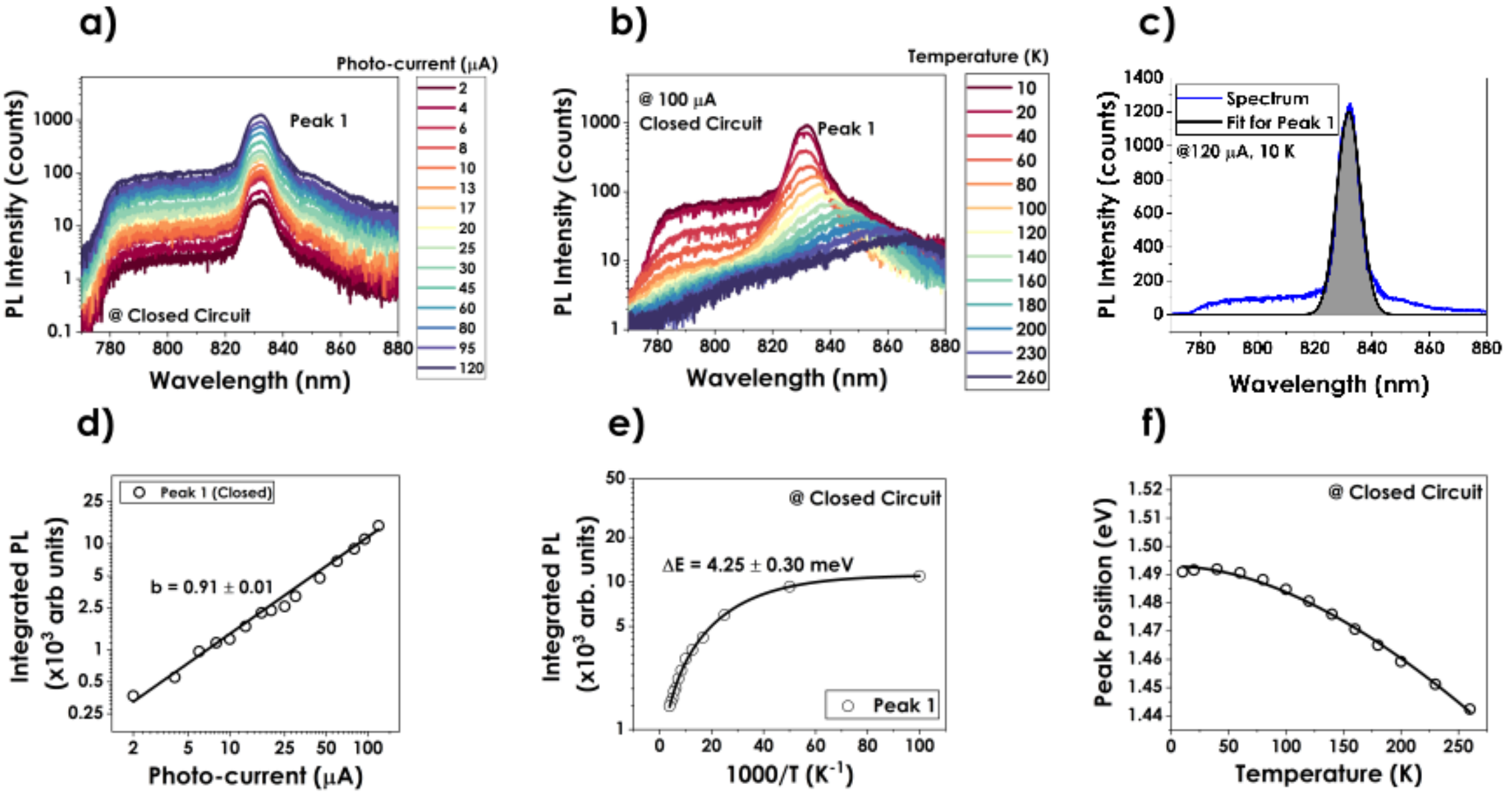

**Supplementary Figure 1.** The figure (a) shows the observed PL spectrum with increasing excitation intensity. (b) shows the observed PL at excitation intensity corresponding to 100$\mu$A with increasing temperature. The plot (c) shows the gaussian fitting of the peak done for the obtained PL to calculate the integrated PL. The plot (d) shows the power law analysis of (a). The plots (e) and (f) show the Arrhenius equation analysis and Varshni analysis of obtained PL (b) with temperature.

Supplementary Figure 1 shows the photoluminescence studies of our sample connected in a closed circuit without applying any external voltage bias. The Supplementary Figure 1(a) shows the measured PL with increasing excitation intensity and Supplementary Figure 1(b) shows the PL at excitation intensity corresponding to 100$\mu$A with increasing temperature. The peak position and integrated PL are estimated from these PL spectra by fitting a gaussian peak of form $y = Ae^{-B(x-C)^2}$ excluding the contributions from the high energy shoulder as the high shoulder corresponds to contributions from substrate as shown in Supplementary Figure 1(c). Form the obtained peak positions of Supplementary Figure 1(a), a power function of form $y = ax^b$ is fitted to show the excitonic nature (b = 1) of the observed peak as shown in Supplementary Figure 1(d). The obtained exponent from the fit for this peak is 1.06 $\pm$ 0.03 which shows that it is excitonic in origin. The integrated PL and peak position of Supplementary Figure 1(b) is then fitted with Arrhenius equation and Varshni equation respectively as shown in Supplementary Figure 1(e) and 1(f). The Arrhenius equation [3,4] is given as $I = A_2 + \frac{A_1 - A_2}{1+\gamma \exp(-\frac{\Delta E}{k_B T})}$ where $\gamma = \frac{\tau_r}{\tau_{nr_0}}$ represents the ratio of radiative recombination to the non-radiative recombination rate, $\Delta E$ represents the energy required to cross the barrier height required for undergoing non-radiative recombination and the constants $A_1$ and $A_2$ are used for normalizing the plot for a better fit. The Varshni equation [5,6] is given as $E_g(T) = E_{g_0} -$ $\frac{\alpha T^2}{\beta+T}$ where $E_{g_0}$ is the bandgap at 0 K and both $\alpha$ and $\beta$ are Varshni parameters. The parameters obtained after both Arrhenius and Varshni fit is given Supplementary Table 1. These Varshni parameters and the bandgap at 0 K are specific to materials and thus can be used to identify the source of the PL signatures. Upon analyses, we conclude that the peak is from the excitons in GaAs.

| | | Arrhenius parameters | | Varshni parameters | | |
|---|---|---|---|---|---|---|
| | | $\gamma$ | $\Delta E$ (meV) | $E_{g_0}$ (eV) | $\alpha$ ($\times 10^{-4}$ eV/K) | $\beta$ (K) |
| Closed Circuit | Peak 1 | 1.29 ± 0.28 | 4.25 ± 0.30 | 1.493 ± 0.001 | 7.20 ± 2.41 | 683.62 ± 313.12 |
| **GaAs** | **Ref.** [3] | **-** | | **1.515 ± 0.001 (for band edge transitions)** | **10.81 ± 2.75** | **588.9 ± 224** |

**Table 1.** This table shows the obtained parameters from the curve fitting of Varshni equation and the Arrhenius equation to measured peak position and the integrated PL of spectra peaks with temperature. Characteristic parameter values of the Varishini equation for GaAs is mentioned at the end along with the references in bold font. By comparing the Varshni parameters with reference, we conclude that the peak is of GaAs.

### 1.2 PL in case of Open-Circuit Measurements:

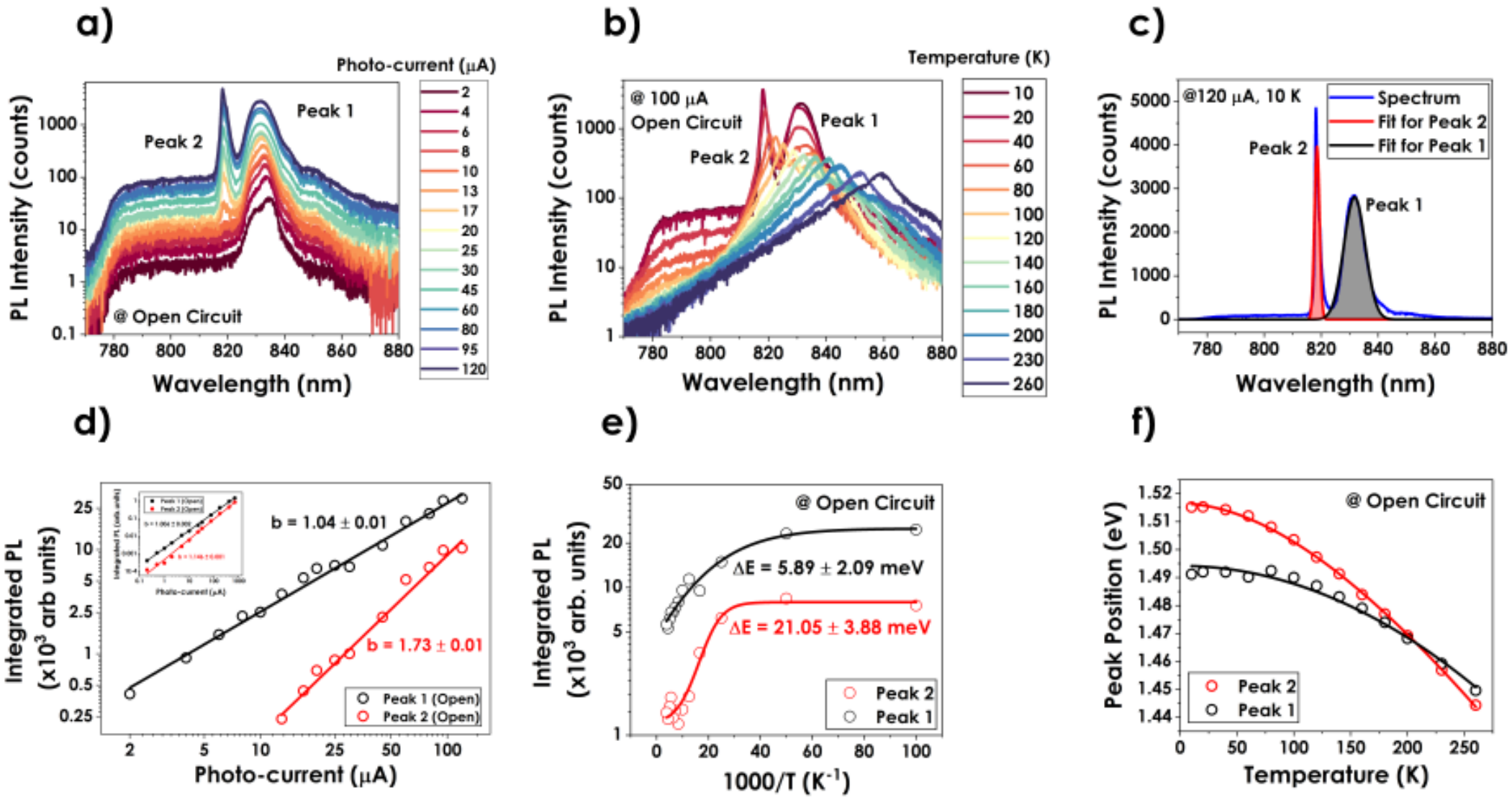

**Supplementary Figure 2.** The figure (a) shows the observed PL spectrum with increasing excitation intensity in case of open circuit. (b) shows the observed PL at excitation intensity corresponding to 100$\mu$A with increasing temperature. The plot (c) shows the gaussian fitting of both the peaks done for the obtained PL to calculate its integrated PL. The plot (d) shows the power law analysis of (a). The plots (e) and (f) show the Arrhenius equation analysis and Varshni analysis of obtained PL (b) with temperature. The peak 2 from the InAs QDs is observed when our sample is not connected in a circuit and the observed sharpness indicates that QDs are very good quality and nearly monodispersed in their size.

Supplementary Figure 2 shows the same photoluminescence studies of our sample when connected in an open circuit. In the spectra we start to see the appearance of the additional peak (peak 2) after certain excitation intensity. Upon analysing the power law dependence of peak 2, in the range of measurement we obtain the exponent of 1.73. But peak 2 survives at higher temperatures, appears at the higher energy side and narrower than the peak 1. Because of these reasons this peak is not of any trion or biexciton, rather should be of exciton from different place than peak 1 as we had also tried to observe the peak 2 at even higher intensities as shown in subplot of Supplementary Figure 2(d). At those higher photo excitations, there we again witness linear dependence indicating the excitonic origin of the same. The temperature analysis of these peaks in open circuit condition shows that the peak 2 should be of InAs QDs as shown in the Supplementary Table 2. Onset of peak 2 at increased excitation intensity is because at such intensity we have more direct excitations in InAs QDs as compared to less excitation intensity. For GaAs, attenuation coefficient $\alpha = 39387\ cm^{-1}$ for photon of 632.8 nm [8] and total thickness travelled to reach photon is 570 nm (approx.) as shown in Supplementary Figure 3(b). The Barrier has higher bandgap thus its absorption is neglected. So, from $I = I_o e^{-\alpha l}$, $I = I_o(0.11)$ (approx.) This shows that Intensity has to increase ~10 times to see peak from InAs QDs as efficient as the peak from GaAs.

| | | Arrhenius parameters | | Varshni parameters | | |
|---|---|---|---|---|---|---|
| | | $\gamma$ | $\Delta E$ (meV) | $E_{g_0}$ (eV) | $\alpha$ ($\times 10^{-4}$ eV/K) | $\beta$ (K) |
| Open-Circuit | Peak 1 | 3.17 ± 3.04 | 5.89 ± 2.09 | 1.494 ± 0.001 | Not a Good Fit to Varshini Equation due to composite nature of peak | |
| | Peak 2 | 137.21 ± 128.64 | 21.05 ± 3.88 | 1.5165 ± 0.0004 | 7.88 ± 0.84 | 472.95 ± 77.21 |
| **GaAs** | **Ref.** [3] | **-** | | **1.515 ± 0.001 (for band edge transitions)** | **10.81 ± 2.75** | **588.9 ± 224** |
| **InAs QD** | **Ref.** [6] | **-** | | **1.099 (for 50nm wide and 6nm thick QDs)** | **7.8** | **459** |

**Table 2.** This table shows the obtained parameters from curve fitting for the observed peak 1 and peak 2 in case of open circuit. By comparing the Varshni parameters with references, we conclude that the peak 1 is of GaAs and peak 2 is of InAs QDs. Similarly, the Arrhenius parameters give an idea of the binding energy of the excitonic species in both GaAs TQW and strongly confined InAs QDs.

Supplementary Figure 2(a) where one need a photo excitation with zero bias photo-current of 10 $\mu$A to see the peak 2 which is from InAs QD. As a result, peak 2 is more prominent at even higher excitation intensity. Now these excited charge carriers (electrons) in QDs will now observe only single barrier as compared to the electrons excited in the top GaAs. Upon calculating the tunnelling probability for an energy diagram as shown in Supplementary Figure 3(a),

Excitation wavelength = 632.8 nm = 1.96 eV

Energy of carriers in InAs QD upon excitation, E = 1.96 eV - 1.6(~Bandgap at 10 K) = 0.36 eV

Barrier height, V = 0.335 eV

Now since E>V,

Tunnelling probability [7], $T = \left[1 + \frac{V^2}{4E(E-V)}(\sin(k_2 a))^2\right]^{-1}$ where $k_2 = \frac{\sqrt{2m(E-V)}}{\hbar}$

For InAs $m_e^* = 0.023\, m_e$

$k_2$ = 0.13 x $10^9$ ($m^{-1}$) and $k_2 a$ = 0.208

So, T = 0.88 (approximately)

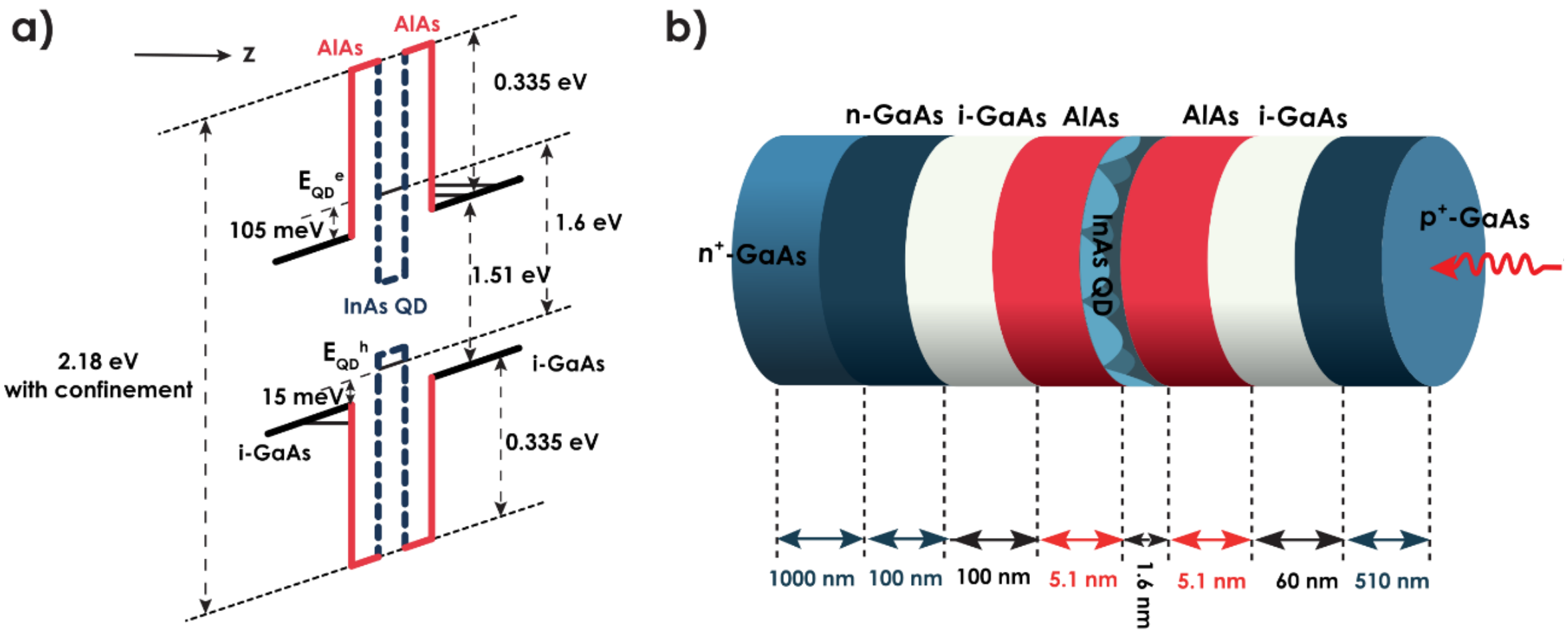

**Supplementary Figure 3.** The figure (a) shows the band structure with appropriate band gaps and energy levels under the condition of reverse bias and the figure (b) shows the thickness of all layers in the sample along with the indication of side of illumination.

It is evident that these electrons have higher probability of tunneling through the barrier. Thus, when these are in closed-circuit there is a creation of channel for the flow of these electrons where some electrons excited at GaAs can also tunnel through the double barrier structure and contribute to dc-photocurrent. So, upon closing the circuit, this leads to a reduction in the availability of the electrons for radiative recombination and thus leading to reduction in the intensity of the emitted PL. However, in case of open-circuit, these electrons have nowhere to flow and recombines inside the InAs QD. As a result, we observe appearance of peak 2 whenever the sample is connected in open-circuit and the same disappears in closed-circuit. To further support this claim, a larger area sample which always has leakage current (a channel for flow of carriers) was studied and observed to not have any PL from InAs QDs. The results are shown in the Supplementary section 3.

**1.3 Analysis of high energy shoulder of PL Spectra:**

To analyze the origin of high energy shoulder observed in the PL spectra, the excitation spot is moved across the sample as shown in Supplementary Figure 4. Upon measuring PL for each such cases, we could see that the shoulder part PL increases as we move away from the sample into the substrate. This shows that the shoulder part is from background and thus neglected at every analysis presented here.

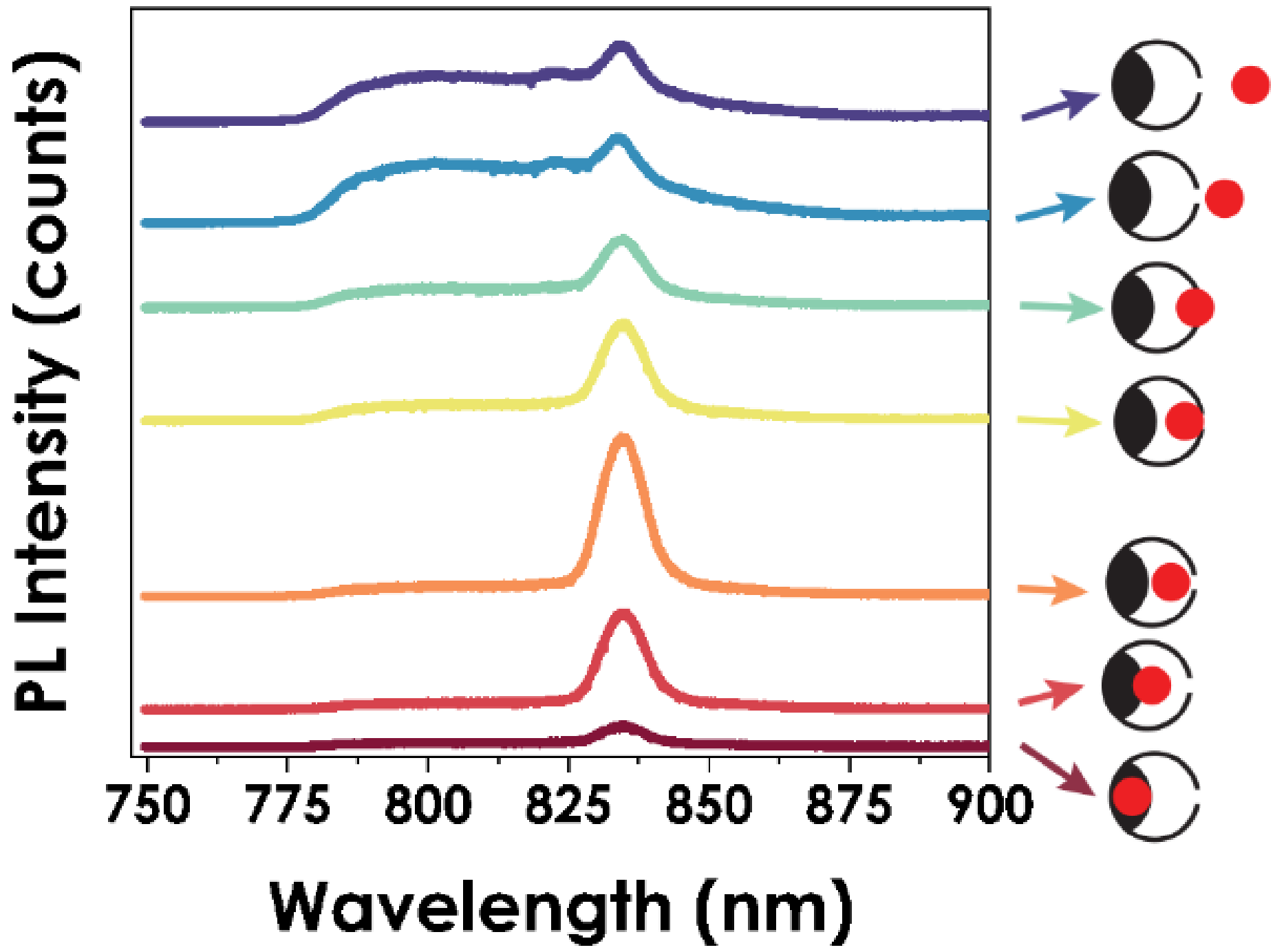

**Supplementary Figure 4.** This plot shows the PL intensity upon moving the excitation spot across the sample. Increase in PL upon spot being away from the sample shows that the high energy shoulder is from the background substrate.

## 2. DC-Photocurrent/Photoluminescence oscillation at higher photo excitation

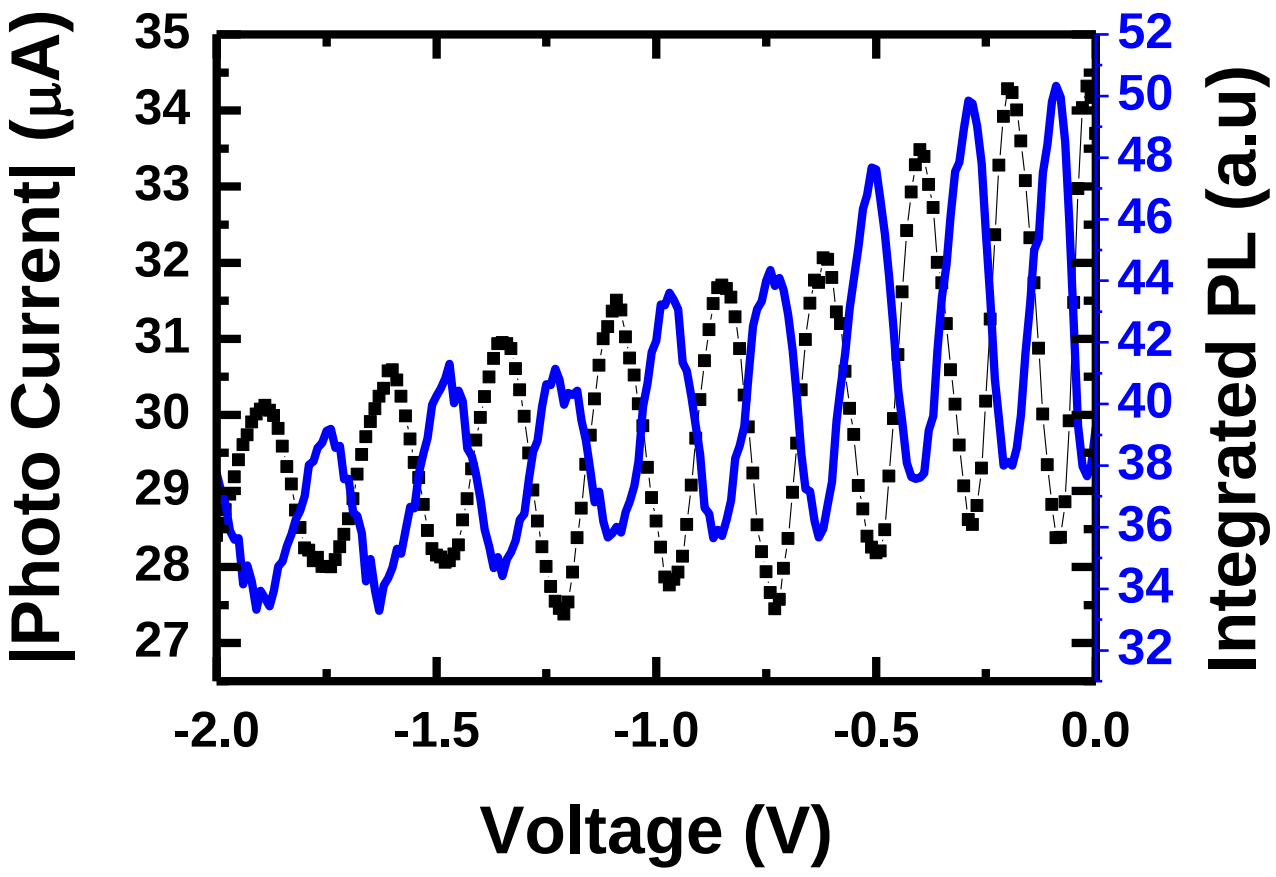

**Supplementary Figure 5.** This plot shows the observed oscillation of integrated PL intensity with applied bias and its 180° out of phase relationship with the absolute magnitude of photo-current oscillation. Data was taken at a higher photoexcitation intensity as compared to those presented in figure 2 of the manuscript.

## 3. Analyses of PL from a sample with bigger ellectrical contact area

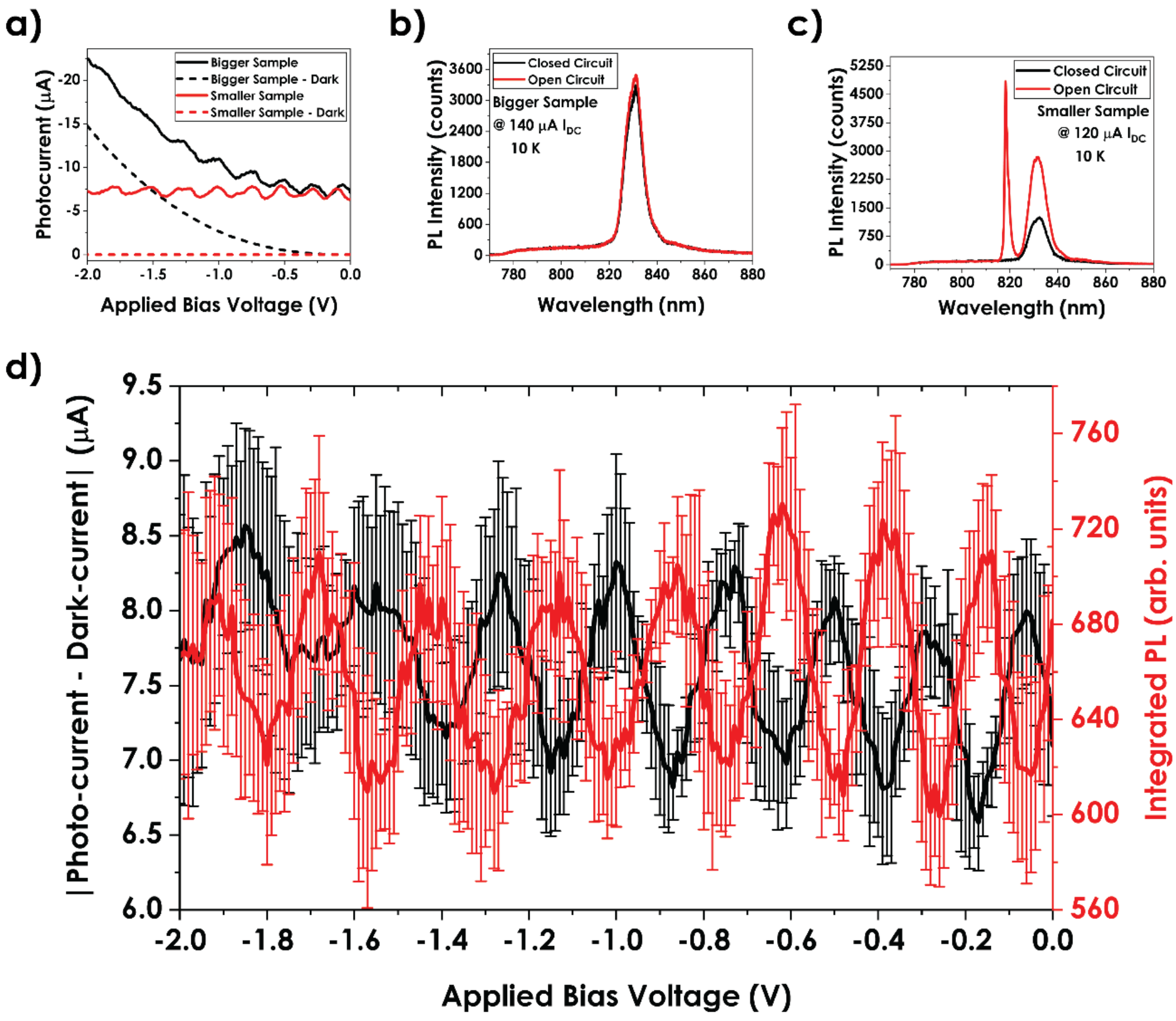

**Supplementary Figure 6.** The figure (a) shows the I-V analysis of both the bigger and smaller sample of 0D-2D heterostructure indicating the expected leaky behaviour of the sample with bigger area. (b) and (c) show a comparison of the PL spectra obtained for cases of closed-circuit and open-circuit conditions in the bigger and smaller samples respectively at higher excitation intensity. The plot (c) shows the PL oscillation observed for the bigger sample.

The Photo-Luminescence (PL) of the same sample with bigger sized (diameter of 400 $\mu$m) electrical contact was also carried out. However, the sample with bigger area turned out to be a leaky diode. This is evident from the I-V analysis of both smaller and bigger sample and shown in Supplementary Figure 6(a). In such a sample, the PL doesn't show any peak 2 even it is excited at higher intensity in case of open-circuit as shown in Supplementary Figure 6(b). In fact, the difference between the case of open-circuit and closed-circuit case in bigger sample is much lesser compared to the difference in case of smaller sample as seen in Supplementary Figure 6(c). It is likely that the bigger sample has 'leaky' barriers leading to darker current at higher bias voltages. So, this leads to less accumulation of charge carriers (electrons) in InAs QDs layer in the bigger sample as there is always a channel for the flow of carriers in this case where electrons can leak through the AlAs barriers Thus the PL peak 2 originated from InAs QDs is not significant. This also supports our argument on why peak 2 actually vanishes upon connecting them in closed-circuit in smaller samples. But this leaking of electrons from barriers doesn't affect the photoluminescence generated in top GaAs layer as the recombination happens over a layer thicker than the QD layer. The observed voltage modulated PL oscillation is also shown in Supplementary Figure 6(d). Despite the mean of the dc-photocurrent oscillation increase with the applied voltage, such effect is not seen in the PL oscillation. This shows that the modulation of PL is associated with the resonant tunnelling of the electrons rather than the photo-current itself.

### 4. Raw Data of AC- AND DC-CONDUCTANCE under dark and under illumination

The Supplementary Figure 7(a) and 7(b) shows the AC and DC conductance measured as a function of bias voltage under dark and illumination condition respectively. This shows that the $G_{AC}^{dark}$ ~ 0 and the $G_{DC}^{dark}$ << $G_{DC}^{ph}$ where the superscript 'dark' referring to no illumination condition and 'ph' referring to under illumination condition. The Fig. 3(d) shows the difference of these two signals, $G_{AC/DC}^{ph} - G_{AC/DC}^{dark}$.

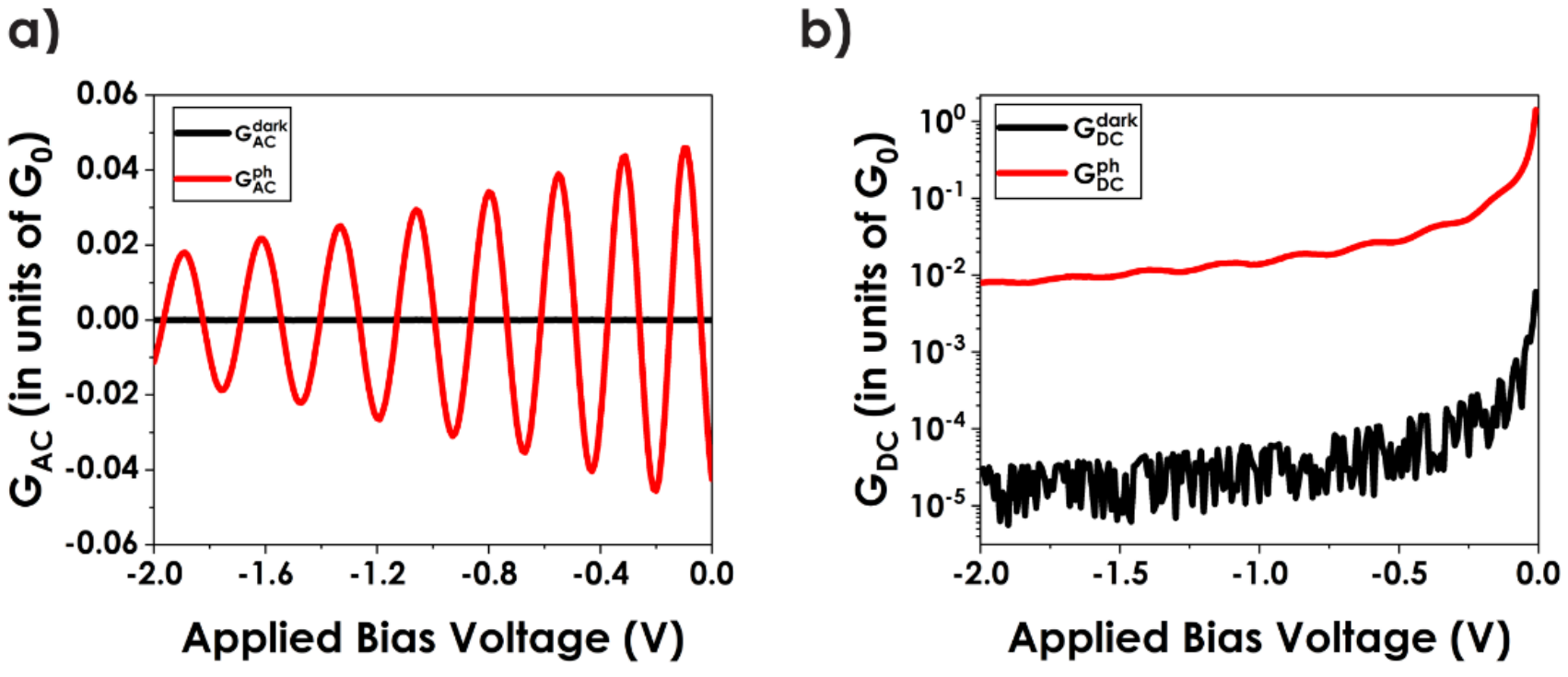

**Supplementary Figure 7.** The figure (a) shows the AC-Conductance measured (at 10kHz frequency and 30mV AC $V_{RMS}$) under dark (black) and illumination (red) conditions and the figure (b) shows the DC-conductance measured under similar conditions with applied reverse bias voltage.

**Supplementary References:**

[1] Datta S and Marie X 2024 *MRS Bulletin* **49** 1

[2] Bhunia A, Singh M K, Huwayz M A, Henini M and Datta S 2023 *Materials Today Electronics* **4** 100039

[3] Datta S, Arora B M and Kumar S 2000 *Phys. Rev. B* **62**, 13604

[4] Fang Y, Wang L, Sun Q, Lu T, Deng Z, Ma Z, Jiang Y, Jia H, Wang W, Zhou J and Chen H 2015 *Scientific Reports* **5** 12718

[5] Varshni Y P 1967 *Physica* **34** 149

[6] Abdellatif M H, Song J D, Lee D and Jang Y 2016 *Applied Science and Convergence Technology* **25** 158

[7] Davis J H 1998 *The Physics of Low Dimensional Semiconductors* (Cambridge University Press)

[8] Aspnes D E, Kelso S M, Logan R A and Bhat R 1986 *J. Appl. Phys.* **60** 754